\documentclass[12pt]{article}
\newcommand{\be}[3]{\begin{equation}  \hlabel{#1#2#3}}     
\newcommand{\ee}{ \end{equation}}
\newcommand{\ba}{\begin{array}}
\newcommand{\ea}{\end{array}}
\newcommand{\NP}[3]{{\em Nucl. Phys.}{ \bf B#1#2#3}}

\renewcommand{\arraystretch}{1.8}

\newcommand{\hjump}[2]{{#2}}
\newcommand{\setref}[1]{ }
\newcommand{\href}[1]{{\ref{#1}}}
\newcommand{\hlabel}[1]{\label{#1} 
  }
\newcommand{\hcite}[1]
{\cite{#1}}
\newcommand{\hbibitem}[1]
{\bibitem{#1} }
\newcommand{\hepth}[7]{
{hep-th/#1#2#3#4#5#6#7}}

\begin{document}

\thispagestyle{empty}
\rightline{HUB-EP-96/55}
\rightline{hep-th/9610232}
\rightline{October 1996}
\vspace{1truecm}
\centerline{\bf \large Quantum corrections for $D=4$ black holes and 
  $D=5$ strings}
\vspace{1.2truecm}
\centerline{\bf Klaus Behrndt\footnote{e-mail: 
 behrndt@qft2.physik.hu-berlin.de}}
\vspace{.5truecm}
\centerline{Humboldt-Universit\"at, Institut f\"ur Physik}
\centerline{Invalidenstra\ss e 110, 10115 Berlin}
\centerline{Germany}

\vspace{2.2truecm}


\vspace{.5truecm}

\begin{abstract}
In recent times quantum corrections for $N=2$ black holes in 4
dimensions have been addressed in the framework of double extreme
black hole solutions, which are characterised by constant scalar
fields. In this paper we generalize these solutions to non-constant
scalar fields. This enables us to discuss quantum corrections for
massless black holes and for configurations that are classically
singular. We also discuss the relation to the 5-dimensional magnetic
string solution.
\end{abstract}

\bigskip \bigskip
\newpage

\noindent
{\bf \large 1. Introduction} \bigskip 

\noindent
Recently there has been a significant progress in the understanding of
$N=2$ black holes in 4 dimensions. The starting point was the
observation that the central charge of a (non-singular) black hole has
an extremum on the horizon and is given by the area of the horizon
\hcite{fe/ka}. In addition, the horizon acts as an attractor on the
scalar fields. Independently where we start at infinity the values of
the scalar fields on the horizon are unique and depend only on the
conserved charges. If one further assumes that the scalar fields are
constant everywhere, i.e.\ they are frozen in their fixpoint, one has
a double extreme black hole \hcite{ka/sh}. They are extreme in the
sense that they saturate the BPS bound ($M=|Z|$) and their central
charge (mass) is extremal (minimal). Since for these solutions the
scalar fields are constant they are a good candidate for the
discussion of quantum corrections. After fixing the classical starting
point \hcite{be/ka}, \hcite{ca/lu} this has been done
\hcite{be/ca}. However, for these black holes it is difficult to
investigate the massless case or to look for quantum solutions that
are stable with less than 4 charges\footnote{On the classical level we
need at least 4 charges to have a non-singular horizon of the black
hole.}. In \hcite{be/ca} a less-charged solution was reachable only
after inclusion of additional topological quadratic terms in the
prepotential and the massless limit of double extreme black holes is 
the flat space time. Furthermore constant moduli are not possible for
massless black holes, since they correspond to a vanishing cycle at
certain point in spacetime (massless black hole singularity). Since
these objects are expected to exist at points of symmetry enhancement
it is desirable to have solutions that allows a massless limit. This
is the case for the solution with non-constant moduli, which we are
going to discuss in this paper. We will restrict ourselves on
axion-free solutions.

\medskip

Before we start let us fix our notation (see \hcite{be/ca} and
refs. therein).  The $N=2$ supergravity includes one gravitational,
$n_{v}$ vector and $n_{h}$ hyper multiplets. In what follows we will
neglect the hyper multiplets, assuming that these fields are
constant. The bosonic $N=2$ action is given by 
\be010 
S \sim \int d^{4}x \sqrt{G} \{ R - 2 g_{A\bar{B}} \partial z^A \, \partial
\bar{z}^B - {1 \over 4 } ( \Im {\cal N}_{IJ} F^I \cdot F^J + \Re {\cal
N}_{IJ} F^{I} \cdot {^{\star}F^{J}}) \} 
\ee 
where the gauge field part $F^I \cdot F^J \equiv F^I_{\mu\nu} F^{J \,
\mu \nu}$ and $I,J = 0,1 .... n_v$.  The complex scalar fields of the
vector multiplets $z^{A}$ ($A=1..n_{v}$) parameterize a special
K\"ahler manifold with the metric $g_{A\bar{B}} = \partial_{A}
\partial_{\bar{B}} K(z,\bar{z})$, where $K(z,\bar{z})$ is the K\"ahler
potential.  Both, the gauge field coupling as well as the K\"ahler
potential are given by the holomorphic prepotential $F(X)$
\be012 \ba{l} 
e^{-K} = i (\bar{X}^I F_I - X^I
\bar{F}_{I}) \\ {\cal N}_{IJ} = \bar{F}_{IJ} + 2i {(\Im F_{IL}) (\Im
F_{MJ}) X^{L} X^{M} \over (\Im F_{MN}) X^{M} X^{N}} 
\ea \ee 
with $F_{I} = {\partial F(X) \over \partial X^{I}}$ and $F_{MN} =
{\partial^{2} F(X)\over \partial X^{M} \partial X^{N}}$ (these are not
gauge field components).  The scalar fields $z^{A}$ are defined by
\be014 
z^{A} = {X^{A} \over X^{0}} 
\ee 
and for the prepotential we take the general cubic form 
\be016 F(X) =  {d_{ABC} X^{A}
X^{B} X^{C} \over X^{0}} 
\ee 
with general constant coefficients $d_{ABC}$. In type II
compactification these are the classical intersection numbers of the
Calabi Yau three fold. On the heterotic side these coefficients
parameterize the general cubic part of the prepotential, which
contains quantum corrections.

\medskip

The paper is organized as follow. In the \hjump{02}{next section} we
describe the 4-dimensional solution. In \hjump{03}{section 3} we describe
the decompactification and especially the relation to the 5-dimensional
magnetic string solution. Finally, we  \hjump{04}{summarize} our
results.  In \hjump{05}{the appendix} we show that the solution given in
section 2 solves the equations of motion.

\bigskip \bigskip

\noindent 
{\bf \large 2. The black hole solution} \setref{02}
\bigskip \newline 
\noindent
In this section we are going to discuss a special solution to the
Lagrangian (\href{010}). We want to restrict ourselves on axion-free
solutions. This means that the scalar fields $z^A$ and as consequence
also ${\cal N}_{IJ}$ are pure imaginary and we will furthermore
simplify the notation in setting $\Im {\cal N}_{IJ} \equiv {\cal N}_{IJ}$.
Then equations of motion are 
\be020
 \ba{c}
 R_{\mu\nu} - 2 \, g_{AB} (\partial_{\mu} z^A
 \partial_{\nu} z^B ) - {1 \over 2} ((F \cdot F)_{\mu\nu} - {1 \over 4} 
 (F \cdot F) G_{\mu\nu} ) = 0 \\
 \partial_{\mu} (\sqrt{G} {\cal N}_{IJ} F^{J \, \mu\nu})) =0 \\
 {4 \over \sqrt{G}} \partial_{\mu} (\sqrt{G} G^{\mu\nu} g_{A\bar B} 
 \partial_{\nu} \bar z^{B}) - 2 (\partial_{A} g_{B\bar C}) 
 \partial z^{B} \,\partial \bar z^{C} - 
 {1 \over 4} (\partial_{A} {\cal N}_{IJ}) F^{I} \cdot F^{J} = 0
\ea
\ee 
where $\partial_{A} = {\partial \over \partial z^{A}}$.

The solution is given in terms of $n_v+1$ harmonic functions $H^A$ and $H_0$
\be030
 \ba{l}
 ds^2 = - e^{-2 U} dt^2 + e^{2 U} d\vec{x} d\vec{x} \qquad , \qquad
 e^{2U} = \sqrt{H_0 \, d_{ABC} H^A H^B H^C} \\
 F^A_{mn} = \epsilon_{mnp}\partial_p H^A \quad , \quad 
 F_{0\; 0m} = \partial_m (H_0)^{-1} \quad , \quad  
 z^A = i H_0 H^A e^{-2U} \ .
\ea
\ee
To be specific we choose for the harmonic functions
\be120
H^{A} = \sqrt{2}( h^{A} + {  p^{A} \over r} ) \qquad , \qquad 
H_{0} = \sqrt{2}( h_{0} + { q_{0} \over r } )
\ee
where $h^{A}$, $h_{0}$ are constant and related to the
scalar fields at infinity. The symplectic coordinates are given by
\be121
 X^0 = e^U \qquad , \qquad 
 X^A = i \, H^A H_0 e^{-U} \ .
\ee
In comparison to the double extreme black holes \hcite{be/ca}, 
every charge has been replaced by a harmonic function.

\medskip

Let us discuss now the charges and the mass of this solution.
The electric and magnetic charges are given by the integrals
over the gauge fields at spatial infinity
\be122
 \ba{l}
q_I = \int_{S^2_{\infty}} {\cal N}_{IJ} {^*F^J} =
  \int_{S^2_{\infty}} {\cal N}_{I0} {^*F^0}  \\
p^I = \int_{S^2_{\infty}} F^J =
 \int_{S^2_{\infty}} F^A \ .
\ea
\ee
In the appendix we show that ${\cal N}_{IJ}$ is diagonal (see (\href{050})).
Thus, the black hole couples to $n_v$ magnetic gauge fields $F^{A}_{mn}$ 
and one electric gauge field $F^0_{0m}$. 

To get the mass we have to look on the asymptotic geometry. 
First, in order to have asymptotically a Minkowski space we have the
constraint $ 4 h_{0} d_{ABC} h^{A} h^{B} h^{C} = 1$. Then
\be130
 e^{-2U} = 1 - {2 M \over r} \pm ....
\ee
Thus we get for the mass
\be140
 M = {q_{0} \over 4 h_{0}}  + 3 p^{A} h_{0} d_{ABC} h^{B} h^{C} \ .
\ee
Using (\href{121}) and calculating the
central charge $|Z|$ we find that the black hole saturates the BPS bound
\be141
M^2 = |Z|^2_{\infty} = e^K_{\infty} (q_0 X^0 - p^A F_A)^2_{\infty}
\ee
where the r.h.s.\ has to be calculated at spatial 
infinity ($e^U_{\infty} =1$).

On the other side if we approach the horizon all constants $h^{A}$ and
$h_{0}$ drop out. The area of the horizon depends only on the conserved
charges $q_0,p^A$. Furthermore, if $q_{0} d_{ABC} p^{A} p^{B} p^{C}>0$ the
solution behaves smooth on the horizon and we find for the area and entropy
(${\cal S}$)
\be150
 A = 4 \, {\cal S} = 4 \pi \sqrt{4 q_{0} \, d_{ABC} p^{A} p^{B} p^{C}}  \ .
\ee
If the charges and $h's$ are positive the area of the horizon defines
a lower bound for the mass.  Minimizing the mass with respect to $h^A$
and $h_0$ gives us the area of the horizon \hcite{fe/ka}
\be170
4 \pi M^{2}|_{min.} = A
\ee
In this case all scalar fields are constant and
\be160
h_{0} = {q_{0} \over c}  \qquad , \qquad h^{A} = {p^{A} \over c}
\ee
where $c^{4} = 4\, q_{0}d_{ABC} p^{A} p^{B} p^{C}$. For these moduli 
all scalars
are constant, i.e. coincides with their value on the horizon ($z^{A} \equiv
z^{A} |_{hor.}$). By this procedure we get the double extreme black holes
\hcite{ka/sh}. Taking this limit our solution (\href{030}) coincides with
the solution given in \hcite{be/ca}. There is yet another way to look on
this extremization. The moduli fields are dynamical fields in $N=2$
supersymmetric gauge theories. Especially the values at infinity ($h^A ,
h_0$) are not protected by a gauge symmetry. Only the electric and magnetic
charges are preserved. For a given model there is no way to fix these
values. Instead one could argue that the model chooses those values for which
the energy or ADM mass is minimal, i.e.\ the double extreme case. This is
the notion of dynamical relaxation that has been introduced in \hcite{re}.
Sofar we have assumed that the charges and modulis are positive.
What happens if some of them are negative? There seems to be no
reason to forbid negative charges. Regarded as compactification of
intersecting branes it simply means that some of them are anti-branes.
Immediately we come to the point of massless black holes \hcite{be}. In the
$N=4$ embedding they correspond on the type II side to a vanishing 2-cycle
in the $K3$. What happens here? Let us look on a simple
example with 3 vector multiplets ($A,B= 1,2,3$) and $d_{ABC}$ given by
\be172
d_{ABC} H^{A} H^{B} H^{C} = H^{1} H^{2} H^{3} + a (H^{3})^{3} 
\ee
with $a>0$.  On heterotic side the first term is the classical $STU$
model and the second term corresponds to a quantum correction to this
model.  Let us regard this as a toy model for discussing the influence
of the quantum corrections on the different types of singularities.

To get massless black holes we can take two charges negative, e.g.\
$p^1$ and $p^2$ (this leaves the entropy invariant). Inserting
these charge into (\href{140}) we find for certain values of
$q_{0}$ or $h^{0}$ massless configuration. For
$e^{4 U}$ we find
\be174
 e^{4 U} = H_0 d_{ABC} H^A H^B H^C = 4 (h_0 + {q_0 \over r}) \left(
 (h^1 - {p^1 \over r})  (h^2 - {p^2 \over r})
 (h^3 + {p^3 \over r}) + a (h^3 + {p^3 \over r})^3 \right) \ .
\ee
The classical solution ($a=0$) discussed in \hcite{be} ($N=4$ embedding)
was pure electric corresponding to $p^2=p^3=q_0 + p^1=0$ and $h_0=h^A=1$ 
(it defines a self-dual case) \footnote{Note, on the heterotic
side $p^1$ becomes electric (one has to go into the ``stringy'' basis).}. 
The general dyonic solution was first given in \hcite{cv/yo}. Obviously 
there is an additional singularity at $r=p^1$.  Classically this is a naked
singularity, it makes the black hole repulsive to all matter
\hcite{ka/li}. In the internal space this singularity corresponds to a
vanishing cycle (see next section).  If we include the quantum
corrections we see that this kind of singularity (at $r= p^1 h^1$ and
$r=p^2 h^2$) vanish. The quantum correction ($\sim a$) acts as a
regulator, not only for the metric (or $e^{4U}$) but also for the
prepotential and Kahler potential (see (\href{121})). To be more concrete,
for positivity the harmonic functions have to fulfill for all radii 
relations like
\be175
H^{1} H^{2} + b \, (H^{3})^{2} > 0
\ee
where $b=a$ for positivity of $e^{4U}$.
For the Kahler metric 
\be177
g_{A\bar{B}} = {3 \over 4 H_{0}} \left( -2 d_{ABC} H^{C} +
3 {(d_{ACD} H^{C} H^{D}) (d_{BEF} H^{E} H^{F}) \over 
d_{ABC} H^{A} H^{B} H^{C}  }\right)
\ee
$g_{33}$ is positive if (\href{175}) holds for $b=a-{2 \over 3}$ and for
$g_{12}$ we have $b=-2a$ (all other components are positive without
restrictions). As long as $a> {2 \over 3}$ we find $p^{1} p^{2} = 2a
(p^{3})^{2}$ and $h^{3}$ has to be large enough (assuming that in (\href{174})
all constants are positive). In the case that $a< {2 \over 3}$ we have at
least for one component of the Kahler metric (e.g. $g_{12}$) a region in space
time where it becomes negative.

What about the horizon at $r=0$? Classically we need all 4 charges in
order to have a non-singular geometry near the horizon.  Also the
double extreme black hole for the cubic prepotential (\href{016}) is
not well defined for less than 4 charges (since the scalar fields
vanish identically).  For this simple quantum model, however, we see
that we can set $p^1=p^2=0$ and still the solution remains
non-singular.  It is still a Bertotti-Robertson geometry with the area
of the horizon
\be176
A = 4 \pi \sqrt{4 a \, q_{0} (p^{3})^{3}} \ .
\ee
and also the scalar fields are non-vanishing.
On the type II side, by including of additional topological terms in the
prepotential, we can even replace the electric charge $q_0$ \cite{be/ca}:
$H_0 = {c_2 \cdot J_3 \over 24} H^3$ and get a non-singular black hole
solution with only one magnetic charge $p^3$ ($c_2$ is the second
Chern class and $J_3$ is a (1,1)-form of the CY three fold).

Sofar we have regarded the model (\href{172}) as a toy model.  To be
realistic we have to discuss the validity of our solution.  The
prepotential (\href{016}) contains only the cubic terms. In general there
are many other terms too. E.g.\ on the type II side we have neglected
all instanton contributions \hcite{ca/os} and on the heterotic side
further quantum corrections \hcite{ha/mo}.  Our
approximation is justified as long as $|z^A| \gg 1$ and for the model
(\href{172}) we have in addition the constraint that
$|z^1|>|z^2|>|z^3|$ (see e.g.\ \hcite{be/ca}). In all spacetime
regions where these inequalities hold our solution is a good
approximation. But we see already that for massless black holes or for
black holes with less than 4 charges we have regions where these
inequalities do not hold.  E.g.\ the classical massless black hole
singularity was given by $H_1=0$ and thus $z^1=0$, so that near this
point our solution is questionable. On the other side we hope
that the incorporation, e.g., of the instanton corrections do
not spoil our statement that the solution behaves regular there.
This, however, needs further investigation.

Since $|z^A| \gg 1$ or $|H_0| \gg 1$ is the decompactification limes
to 5 dimensions let us look what the 5 dimensional solution looks
like.

\bigskip \bigskip

\noindent
{\bf \large 3. Decompactification} \setref{03}
 \bigskip  \newline
\noindent
There are many ways to get the solution (\href{030}) by
compactification of higher-dimensional configurations. On the
heterotic side it is a compactification on $K3 \times
T_2$. Classically our solution corresponds to the 6-dimensinal
solution discussed in \hcite{cv/ts}.  On the type II side we have a CY
compactification, e.g.\ of three $D$-4-branes and a $D$-0-brane for
type IIA string theory. Alternatively we can see our solution as a
compactification of an intersection of 3 $M$-5-branes \hcite{pa/to}
and a boost
along the common the string.  Let us discuss the last possibility in
more detail. If we have only $C_{123}=1$ our solution (with 3 moduli
$A=1,2,3$) corresponds to the following intersection in 11 dimensions
\hcite{ts}
\be220
ds^2_{11} =  {1 \over (H^1 H^2 H^3)^{1 \over 3}} \left[ du dv + H_0 du^2
 + H^1 H^2 H^3 d\vec{x}^2 + H^A \omega_A \right] \ .
\ee
This is a configuration where three 5-branes intersect over a common string
and each pair of 5-branes intersect over a 3-brane. In going to 4 dimensions
we first compactify over $H^A \omega_A$, with $\omega_A$ defining three
2-dimensional line elements. After this we are in 5 dimensions and have a
string solution with momentum modes parametrized by $H_0$ ($H^A$ are
parameterizing the 5-branes). Before we generalize this solution let us look
once more in the massless black hole singularity, which was given, e.g.\ ,
by $H^1 =0$. At this point we see that one of the 2-cycles vanish ($ H^1
\omega_1 =0$). Thus, classically there is not only 
a singularity in the 4-dimensional
spacetime but also in the internal space. Generalizing this solution to a
generic CY with non-trivial intersection numbers we find for this
5-dimensional string solution
\be230
ds^2 =  {1 \over (d_{ABC} H^A H^B H^C)^{1 \over 3}} \left( dv du  + H_0 du^2 
 + (d_{ABC}H^A H^B H^C ) d\vec{x}^2 \right) \ .
\ee
Compactifying this string solution over $u$ yields our 4-dimensional
black hole solution (\href{030}). The electric gauge field is a
Kalluza-Klein field, in 5 dimensions we have only magnetic gauge
fields which are the same as in $D=4$. In addition one of the
4-dimensional scalar fields is the compactification radius, which is
related to $|H_0|$ and thus $|H_0|\gg 1$ gives us the
decompactification limes, for which our solution is good
approximation.

For the generic case ($d_{ABC} p^A p^B p^C \neq 0$) this 5-dimensional string
solution is non-singular and the asymptotic geometry near the horizon is
given by $AdS_3 \times S_2$. 

\bigskip

\bigskip \bigskip

\noindent
{\bf \large 4. Discussion} \setref{04} \bigskip 

\noindent
In this paper we have generalized the $N=2$ double extreme black hole
solution in \hcite{be/ca} to the case of non-constant scalar fields (see
(\href{030})). This solution allows also a massless limit. In the classical
limit the massless solution has a naked singularity where one cycle of the
internal space vanishes. We argued that a cubic correction term
(\href{172}) can act as an regulator for this singularity. Also we saw
that for this model we can turn off 2 charges and still have a non-singular
horizon. On the other side we have to take these results with care, since this
model is only a good approximation for large moduli. E.g.\ near the massless
black hole singularity this is not the case. 

In a second part we have discussed the 5-dimensional magnetic string
solution that yields upon compactification this black hole in 4 dimensions.
This connection could be interesting with respect to the $D$-brane
picture and microscopic state counting. The similarity to the
state counting for 5-dimensional black holes in \hcite{st/va}
is obvious. It should be possible to repeat their calculation,
but now with $K3$ replaced by the CY three fold and the intersection
numbers by $d_{ABC} p^A p^B p^C$. Again the electric charge is
related to the momentum modes travelling along the string.
We hope to come back to this point in the future.

\bigskip  \bigskip
\newpage

  \setref{05} \noindent
{ \bf  \large A. The field equations}
\bigskip

\noindent
Let us convince that the solution (\href{030}) really solves the above
equations (\href{020}). First, we know that if we have only $d_{ABC} =
d_{123}$ it is a solution. These are the known classical black hole
solutions. It is also a solution if $H^A \sim p^A f$ (where $f$ is a
harmonic function) which is the double extreme black hole. We have to
discuss the case for arbitrary symmetric $d_{ABC}$ and
for general harmonic functions. In addition, for this discussion the
function $H_0$ is irrelevant. If we are sure that it is a solution for
$H_0=1$ we can decompactify this solution to $D=5$ and can generate
the $H_0$ function by a boost along the 5d (magnetic) string.  (see
(\href{230})).

\medskip

\noindent
{\bf Gauge field equation}

\smallskip \noindent
For this we need an expression for ${\cal N}$. Our prepotential
and their derivative is given by
\be040
 \ba{l}
 F = {d_{ABC} X^A X^B X^C \over X^0} \\
 F_0 = - X^0 \, d_{ABC} z^A z^B z^C  \qquad , \qquad 
 F_A = 3 X^0 \, d_{ABC} z^B z^C \\
 F_{00} = 2\, d_{ABC} z^A z^B z^C  \quad  ,\quad 
 F_{0A} = -3 \, d_{ABC} z^B z^C \quad , \quad F_{AB} = 6 \, d_{ABC} z^C 
\ea  
\ee
where $z^A = {X^A \over X^0}$. Note, that since $z^A$ is imaginary
(axion-free) $F_{0A}$ is real and all other second derivatives are imaginary.
Inserting these terms now into (\href{012}) we find
\be050
 {\cal N}_{00} = - d_{ABC}z^A z^B z^C \  , \  {\cal N}_{0A} = 0 \ ,
 \ {\cal N}_{AB} = -6 (d_{ABC} \, z^C) + 9 \, {(d_{ADE} \, z^D z^E )
 (d_{BCF} \, z^C z^F ) \over (d_{ABC} z^A z^B z^C )}  \ .
\ee
As consequence, we have $\Re {\cal N}=0$. Inserting this into the
electric field equations we find
\be060
 \partial_m \left(\sqrt{G} {\cal N}_{00} F^{0 \; m0}) = 
 \partial_m ((H_0)^2 \partial_m {1 \over H_0}\right) = 0
\ee
since $H_0$ is harmonic.. The magnetic field equations are solved
due to the ansatz (lets ignore here the subtleties
with multi-center solutions). And the Bianchi identities are
solved for harmonic $H^A$.

\medskip

\noindent
{\bf Einstein equations}
\smallskip

\noindent
For Ricci tensor we have
\be070
R_{mn} = - \partial^2 U \; \delta_{mn} - 2 \partial_m U \partial_n U
\qquad , \qquad R_{00} = - \partial^2 U \ e^{-4 U} 
\ee
where $\partial^2 = \delta_{mn} \partial_m \partial_n$. Let us now set
$H_0=1$. As next step we calculate the Kahler metric. Taking $K= - \log (-i
d_{ABC}(z-\bar{z})^A(z-\bar{z})^B(z-\bar{z})^C)$ we find ($\bar z^A = -z^A$)
\be080
g_{A\bar{B}} = {1 \over 4} \left( 6 \, {(d_{ABC} z^C) \over 
(d_{ABC} z^A z^B z^C)} -
9 \, {(d_{ADE} z^D z^E) (d_{BCF} z^C z^F) \over (d_{ABC} z^A z^B z^C)^2} 
\right) \ .
\ee
For $(F \cdot F)_{mn}$ we find
\be100
\ba{rl}
 {\cal N}_{AB} F^A_{mp} F^B_{nq} G^{pq} & =  e^{-4U} \left(
( - \partial^2 {\cal D} + {\partial D \partial D \over {\cal D}} )  \; 
 \delta_{mn}  +\partial_m \partial_n {\cal D} - {\partial_m{\cal D} 
 \partial_n{\cal D}  \over {\cal D}}  
  -3 \, D_A \partial_m \partial_n H^A \right) \\
 & = - 4 \left( \partial^2 U \, \delta_{mn} - \partial_m \partial_n U \right)
  -3 D_A \partial_m \partial_n H^A 
\ea
\ee
where we are using the compact notation ${\cal D} = d_{ABC} H^{A} H^{B}
H^{C}$, ${\cal D}_A = d_{ABC} H^{A} H^{B}$. The scalar field part yields
\be110
 \ba{rl}
 g_{A\bar{B}} \partial_m z^A \partial_n \bar{z}^{B} & =  
  {1 \over 4} e^{-4 U} \left( - \partial_m \partial_n{\cal D} 
  + {\partial_m {\cal D} \partial_n {\cal D} \over {\cal D}}
  +3 \, {\cal D}_A \partial_m \partial_n H^A \right) \\
 & =  - \partial_m \partial_n U - \partial_m U \partial_n U + {3 \over 4}
\,  {\cal D}_A \partial_m \partial_n H^A
\ea
\ee
If we now insert these expression into (\ref{020})  we
find that the Einstein equations are fulfilled.
 
\medskip 

\noindent
{\bf Scalar field equations}

\smallskip \noindent
This equation consists of three terms. Let us start with the first
one. 
\be112
\ba{l}
 {4 \over \sqrt{G}} \partial_{\mu} (\sqrt{G} G^{\mu\nu}  
 g_{A\bar B}\partial_{\nu} \bar z^{B}) = \\ \qquad = 3i \, e^{-4 U} \left(
 \partial^2 {\cal D}_A - {1 \over 2} e^{-4U} {\cal D}_A \partial^2 
 {\cal D}  - e^{-4U} \partial {\cal D}_A \partial {\cal D} +
 {3 \over 4} e^{-8U} {\cal D}_A \partial {\cal D} \partial {\cal D} \right)
\ea
\ee
Next we find
\be114
 2 (\partial_{A} g_{B\bar C}) \partial z^{B} \partial \bar z^{C} = 
 {3 i \over 2} e^{-4U} \left( \partial^2 {\cal D}_A - e^{-4U} {\cal D}_A
 \partial^2 {\cal D} + {1 \over 2} e^{-8U} {\cal D}_A \partial {\cal D}
 \partial {\cal D} \right) \ .
\ee
Finally for the gauge field part we have
\be116
 {1 \over 4} \partial_{A} {\cal N}_{BC} \, F^{B} \cdot F^{C} =
 {3 i \over 2} e^{-4 U}  \left( \partial^{2} {\cal D}_A  - 2 e^{-4U} 
 \partial {\cal D}_A \partial {\cal D} + e^{-8U} {\cal D}_A \partial {\cal D}
 \partial {\cal D} \right) \ .
\ee
Putting all these expressions into (\ref{020}) we find that also the
scalar field equation is fulfilled.

\bigskip \bigskip
\bigskip

\noindent
{\bf Acknowledgements}  \medskip \newline
The work is  supported by the DFG. I would like to thank R.\ Kallosh,
T.\ Mohaupt and G.\ Behrndt for useful comments and many discussions.
In addition I am grateful to K.Z.\ Win for drawing  my attention
to some typos in an earlier version of the file.

\newpage


\renewcommand{\arraystretch}{1}

\end{document}